\documentclass[journal,10pt]{IEEEtran}

\usepackage{amssymb,amsmath,times,comment}
\usepackage{cite}
\usepackage{epsfig}
\usepackage{makecell}
\usepackage{ctable}
\usepackage{tabularx,booktabs}
\usepackage{diagbox}
\usepackage[hyphens]{url}
\usepackage{hyperref}
\usepackage[acronym]{glossaries}
\usepackage{hyperref}
\usepackage{algpseudocode}
\usepackage[ruled,vlined,linesnumbered]{algorithm2e}

\usepackage{amsmath, amssymb, bm} 

\title{Adaptive Pinching Antenna Optimization via Meta-Learning for Physical-Layer Security in Dynamic Wireless Networks}

\author{{Khalid~T.~Musri,~\IEEEmembership{Member,~IEEE},  Akram~Y.~Sarhan,~\IEEEmembership{Member,~IEEE},\\ Osamah~A.~Abdullah,~\IEEEmembership{Member,~IEEE}, and Hayder Al-Hraishawi,~\IEEEmembership{Senior Member,~IEEE}
\thanks{
K. T. Musri is with Department of Cybersecurity, University of Jeddah, Saudi Arabia.\\
A. Y. Sarhan is with the Department of Information Technology, College of Computing and Information Technology, University of Jeddah, Saudi Arabia.\\
O. A. Abullah is with the Ministry of Higher Education and Scientific Research, Baghdad, Iraq.\\
H. Al-Hraishawi is with the Department of Electrical Engineering, University of South Florida, Tampa, FL 33620 USA.\\
Corresponding author: \emph{Akram Y. Sarhan (asarhan@uj.edu.sa)}.}
\thanks{This work was funded by the University of Jeddah, Jeddah, Saudi Arabia, under grant No. (UJ-25-DR-805). Therefore, the authors thank the University of Jeddah for its technical and financial}}}

\begin{document}
\maketitle

\begin{abstract}
This paper develops a gradient-based meta-learning framework for real-time control of waveguided pinching-antenna systems under user-location uncertainty and physical-layer security (PLS) constraints. A probabilistic system model is introduced to capture the impact of imperfect localization on outage performance and secrecy. Based on this model, a joint antenna-positioning and transmit-power optimization problem is formulated to satisfy probabilistic reliability and secrecy requirements. To enable rapid adaptation in highly dynamic environments, the proposed approach employs model-agnostic meta-learning (MAML) to learn a transferable initialization across diverse mobility and channel conditions, allowing few-shot online adaptation using limited pilot feedback. Simulation results demonstrate that the proposed framework significantly outperforms Reptile-based meta-learning, non-meta reinforcement learning, conventional optimization, static antenna placement, and power-only control in terms of outage probability, secrecy performance, and convergence latency. These results establish meta-learning as an effective tool for secure and low-latency control of reconfigurable pinching-antenna systems in non-stationary wireless environments.
\end{abstract}




\section{Introduction}
Emerging sixth-generation (6G) applications, including extended reality (XR), immersive telepresence, the tactile Internet, and autonomous systems, impose stringent requirements on wireless communication links in terms of reliability, latency, energy efficiency, and security \cite{saad2019vision,10753482}. Meeting these demands in practice is particularly challenging in highly dynamic environments, where wireless channels are affected by user mobility, blockage, interference, and localization errors \cite{bennis2018ultrareliable}. These challenges are further exacerbated at millimeter-wave (mmWave) and terahertz (THz) frequencies, especially in ultra-dense deployments, where propagation is highly directional and even small spatial misalignments can lead to severe performance degradation \cite{Rappaport2013,9882323}. To address these limitations and enhance the effective signal-to-noise ratio (SNR), recent research has increasingly focused on reconfigurable antenna architectures that dynamically adapt their physical location or radiation characteristics to changing propagation conditions \cite{Zhu2024}. Such flexible positioning enables mitigation of small-scale fading and blockage effects, resulting in substantial gains in link reliability and spectral efficiency \cite{Zhou2025}.

In this context, \emph{pinching antennas} have recently emerged as a promising realization of flexibly positioned antenna systems \cite{Ding2025}. By “\textit{pinching}” small dielectric elements onto a waveguide, these antennas create controllable radiation points that can be repositioned along the waveguide, enabling highly directive, user-centric line-of-sight (LOS) links over large apertures \cite{suzuki2022pinching}. Unlike conventional fixed-position antennas, pinching antennas dynamically adapt their radiating location, providing an additional spatial degree of freedom that facilitates dynamic beam shaping, enhanced coverage, and reduced interference \cite{Mao2025}. This mechanical reconfigurability enables adaptive and power-efficient connectivity beyond what is achievable with fixed or electronically steered arrays \cite{Samy2025}. Moreover, due to their simple structure and minimal activation overhead, pinching antennas offer improved cost and energy efficiency compared to traditional antenna arrays \cite{11169486}. Despite these advantages, their operation introduces new challenges in adaptive control and system integration, particularly under mobility, uncertainty, and security constraints \cite{Yang2025}.

The same flexibility that makes pinching antennas attractive also introduces daunting operational challenges \cite{Singh2025}. Fully exploiting their capabilities requires intelligent control mechanisms capable of determining the optimal antenna position and transmit power under imperfect channel state information (CSI), uncertain user localization, and rapidly varying propagation conditions. Specifically, user locations are often estimated with non-negligible error due to imperfections in positioning and tracking methods, feedback latency, and user mobility \cite{10540175}, all of which complicate beam alignment and degrade both system performance and link reliability \cite{10287946}. Moreover, the spatial reconfigurability of pinching antennas can create strong unintended side links, increasing vulnerability to eavesdropping if pinch placement and radiation control are not carefully managed \cite{Zhu2025}.

Recent studies on pinching antenna activation have demonstrated that dynamically selecting radiating points can significantly enhance energy efficiency and spectral utilization  \cite{Tyrovolas2025}. For instance,
the works in \cite{10896748} and \cite{Tegos2025} optimized pinching-antenna locations for downlink and uplink communication scenarios, respectively. The study in \cite{10912473} examined the activation of multiple pinching antennas to support multiple users in a non-orthogonal multiple access (NOMA) downlink setting. In \cite{Qin2025}, joint optimization of pinching-antenna placement and user power allocation was proposed to enhance both communication and sensing performance in integrated sensing and communication (ISAC) systems. 

In parallel, several efforts have examined the use of physical-layer security (PLS) techniques in pinching antenna-enabled networks to enhance secrecy performance and mitigate information leakage in the presence of an eavesdropper (Eve) \cite{papanikolaou2025}. For instance, \cite{Wang2025} developed a secrecy-rate maximization framework in which multiple pre-installed pinching antennas  adjust amplitude and phase, modeled through a coalitional game with Shapley-value-based activation, to strengthen the legitimate link while suppressing Eve. Likewise, \cite{11205176} introduced a pinching antenna-enabled index and directional modulation  scheme that integrates index modulation, directional modulation, adaptive antenna repositioning, and artificial-noise (AN) injection to degrade the Eve decoding capability.

However, most existing works formulate antenna activation, beamforming, and PLS provisioning as mixed combinatorial optimization problems and solve them via heuristics, convex relaxations, or iterative algorithms. These approaches mostly rely on assumptions of perfect CSI, quasi-static channels, and stationary user positions, assumptions that break down in mobile environments where channel coherence times are short, estimation errors accumulate, and user motion introduces both uncertainty and rapidly varying propagation conditions. To overcome these limitations, recent studies have begun investigating learning-based adaptation approaches. In particular, a small number of works have explored meta-learning for fast adaptation under non-stationary wireless conditions \cite{zhou2025gradient, amhaz2025coordinated, Zhang2024}. Unlike conventional supervised or reinforcement learning (RL) approaches, meta-learning learns transferable priors across tasks, enabling rapid adaptation to new channel states or mobility patterns from only a few observations \cite{frikha2021few, Hospedales2022}. This property makes meta-learning suitable for environments where real-time reconfiguration is essential and only limited feedback can be obtained at run time.

Deploying PLS in pinching-antenna systems is challenging under mobility and channel uncertainty, especially in adversarial environments \cite{Hayder2017}. Existing works typically address these aspects in isolation: activation-oriented studies focus on optimizing radiation-point placement, whereas PLS-centric approaches emphasize secrecy enhancement through beam steering or AN injection. To the best of our knowledge, no prior research integrates meta-learning with PLS in pinching-antenna systems, nor provides a unified framework that jointly handles real-time adaptation, user-location uncertainty, and adversarial channel dynamics.
These challenges are compounded by the fact that the associated optimization problem is fundamentally non-convex, driven by probabilistic outage constraints, the secrecy-rate coupling between legitimate users and illegitimate users, and the nonlinear dependence of channel gains on user location and antenna position. Moreover, in fast-varying mobile environments, repeatedly solving such computationally intensive problems from scratch is infeasible for real-time operation \cite{8227766}.

Motivated by these limitations, this work proposes a gradient-based meta-learning controller that enables efficient and secure antenna operation under uncertainty. Leveraging both \textit{model-agnostic meta-learning (MAML)} \cite{finn2017model} and \textit{Reptile} \cite{nichol2018first} algorithms, the framework learns a transferable initialization across diverse legitimate user-Eve configurations, allowing the access point (AP) to jointly adjust antenna position and transmit power using only a few pilot measurements. This leads to a highly sample-efficient system capable of real-time secrecy provisioning in non-stationary and uncertainty-loaded wireless conditions \cite{10542646, zhu2024robust}. In brief, the key contributions of this paper are outlined as follows:
\begin{itemize}
    \item We develop a novel framework for pinching-antenna systems that jointly integrates antenna positioning and transmit-power adaptation under PLS constraints, explicitly accounting for user-location uncertainty and the presence of an Eve.

    \item We formulate a secrecy-aware optimization problem that minimizes transmit power subject to probabilistic outage and secrecy-rate constraints, capturing the nonlinear dependence of the channel on antenna position and the intrinsic coupling between legitimate and eavesdropping links.

    \item To overcome the computational intractability of solving the resulting non-convex problem in real time, we propose a gradient-based meta-learning controller based on MAML/Reptile principles, enabling few-shot adaptation of antenna position and transmit power under previously unseen and rapidly varying conditions.

    \item To the best of our knowledge, this is the first work to jointly combine antenna activation and PLS provisioning within a meta-learning framework for pinching-antenna systems, allowing rapid response to mobility, localization uncertainty, and adversarial dynamics.

    \item Extensive simulations demonstrate that the proposed framework outperforms non-meta RL, classical optimization, static antenna positioning, and power-only control in terms of outage probability, secrecy performance, convergence latency, and transmit-power efficiency.
\end{itemize}
These contributions advance pinching-antenna research beyond  static or non-adversarial settings and establish meta-learning as an effective tool for secure, low-latency control in non-stationary wireless networks.

The remainder of this paper is organized as follows. Section~\ref{sec_system_model} presents the system model, while Section~\ref{sec_security_constraints} formulates the secrecy-constrained optimization problem. Section~\ref{sec_proposed_framework} introduces the proposed gradient-based meta-learning framework and the corresponding online adaptation algorithm. Section~\ref{sec_simulations} describes the simulation environment and discusses the numerical results. Finally, Section~\ref{sec_conclusions} concludes the paper.

\noindent\textbf{Notation:}
The operators $\mathbb{E}[\cdot]$ and $\mathbb{P}[\cdot]$ denote expectation and probability, respectively. 
The Euclidean norm and absolute value are denoted by $\|\cdot\|$ and $|\cdot|$, respectively, and $Z \sim \mathcal{CN}(\mu,\sigma^2)$ indicates that $Z$ is a circularly symmetric complex Gaussian random variable with mean $\mu$ and variance $\sigma^2$.

\begin{figure}[t]
    \centering
    \includegraphics[width=1\columnwidth]{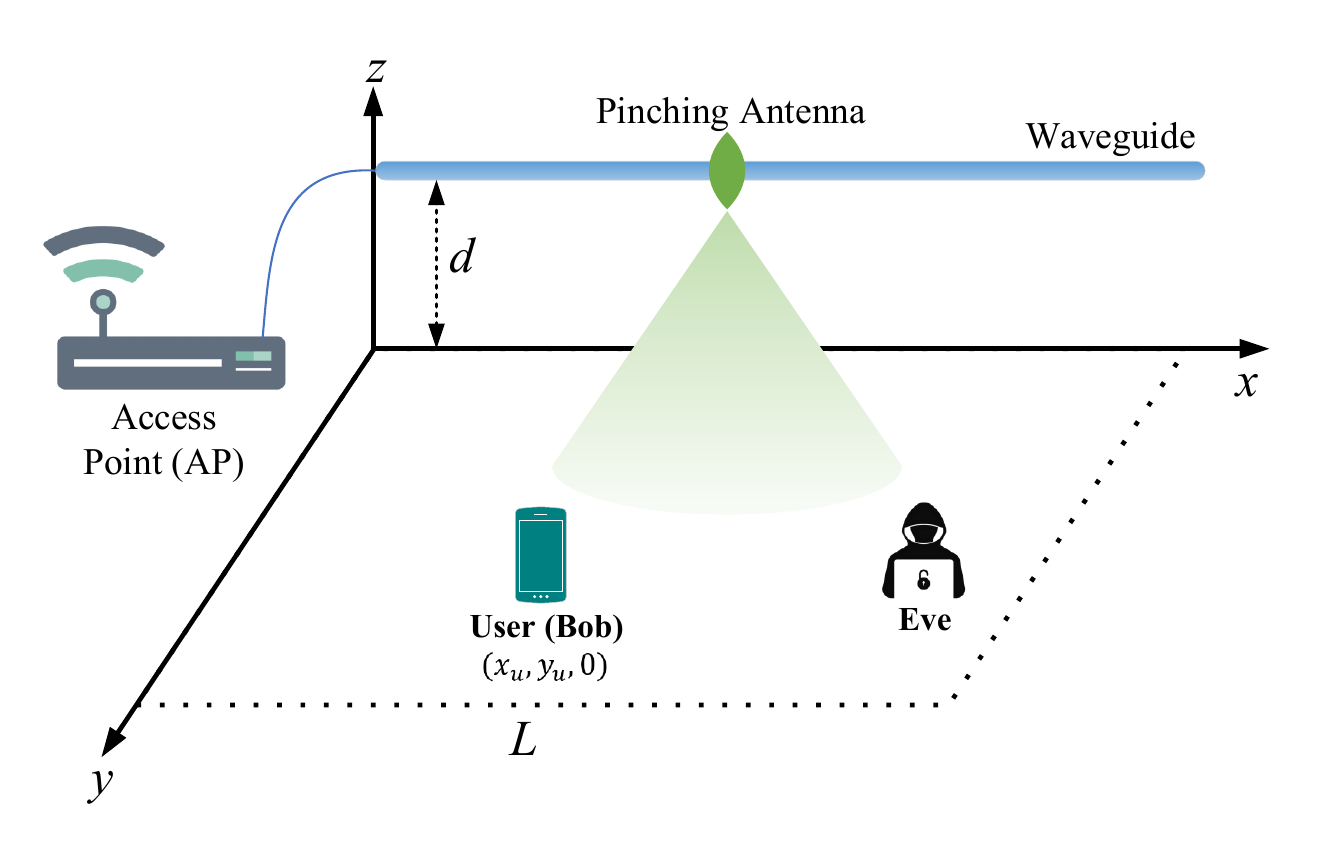}
    \caption{Schematic diagram of a dynamic downlink communication system with a waveguided pinching antenna serving a legitimate user (Bob) in the presence of an eavesdropper (Eve).}
    \label{fig:system_model}
\end{figure}

\section{System Model and Problem Formulation}\label{sec_system_model}

As shown in Fig.~\ref{fig:system_model}, we consider a dynamic downlink communication scenario in which an AP serves a legitimate user (Bob) in the presence of a potential Eve. The AP is equipped with a single pinching antenna mounted on a linear dielectric waveguide of length $L$ meters, elevated at a height $d$ meters above the ground plane. In contrast to conventional fixed-position antennas, the pinching antenna is mechanically reconfigurable and can slide along the waveguide, enabling spatial alignment with the user distribution and mitigation of signal leakage toward unintended receivers. This mechanical agility introduces a hardware-level spatial degree of freedom that complements digital signal processing while maintaining a low-complexity, single RF-chain architecture.
The three-dimensional (3D) position of the pinching antenna is given by
\begin{equation}
\boldsymbol{\Phi}_{\mathrm{PA}} = (x_{\mathrm{PA}}, 0, d), \quad x_{\mathrm{PA}} \in [0, L],
\end{equation}
where $x_{\mathrm{PA}}$ denotes the controllable horizontal position along the waveguide and $d$ is the fixed elevation above the ground. The AP dynamically adjusts $x_{\mathrm{PA}}$ in response to environmental variations such as user mobility and channel fluctuations.

\subsection{User Location Uncertainty}

Users are randomly located in a two-dimensional horizontal area beneath the waveguide. For the legitimate user (Bob), the estimated position is denoted by $\mathbf{u} = (x_u, y_u, 0)$. Due to imperfections in positioning and tracking mechanisms, feedback latency, and user mobility, the actual position $\hat{\mathbf{u}}$ may deviate from this estimate. This uncertainty is modeled by a circular region of radius $r_u$, within which the user is uniformly distributed:
\begin{equation}
\|\hat{\mathbf{u}} - \mathbf{u}\|^2 \leq r_u^2.
\end{equation}
This model captures realistic localization uncertainty in mobile wireless environments and motivates the need for robust and adaptive transmission strategies.

\subsection{Signal and Channel Model}

Let the AP transmit a complex symbol $s \in \mathbb{C}$ with $\mathbb{E}[|s|^2]=1$ at transmit power $P$. The received signal at any spatial location $\mathbf{z} \in \mathbb{R}^3$ (corresponding to Bob or Eve) is
\begin{equation}
y(\mathbf{z}) = \sqrt{P}\, h(\mathbf{z}, \boldsymbol{\Phi}_{\mathrm{PA}})\, s + n(\mathbf{z}),
\label{eq:received_signal}
\end{equation}
where $n(\mathbf{z}) \sim \mathcal{CN}(0,\sigma^2)$ denotes additive white Gaussian noise and $h(\mathbf{z}, \boldsymbol{\Phi}_{\mathrm{PA}})$ is the channel coefficient.

We adopt a dominant LOS free-space propagation model, which is appropriate for mmWave/THz frequencies and aligns with the highly directional nature of pinching antennas. Small-scale fading and shadowing are omitted to isolate the impact of antenna position control. The channel coefficient is expressed as
\begin{equation}
h(\mathbf{z}, \boldsymbol{\Phi}_{\mathrm{PA}}) =
\sqrt{\frac{\lambda^2}{(4\pi)^2 \|\mathbf{z} - \boldsymbol{\Phi}_{\mathrm{PA}}\|^2}}
\, e^{-j \frac{2\pi}{\lambda} \|\mathbf{z} - \boldsymbol{\Phi}_{\mathrm{PA}}\|},
\label{eq:channel_coefficient}
\end{equation}
where $\lambda = c/f_c$ is the wavelength corresponding to carrier frequency $f_c$, and $\|\cdot\|$ denotes the Euclidean norm. This model captures distance-dependent path loss and phase rotation due to propagation delay.

The corresponding SNR at location $\mathbf{z}$ is
\begin{equation}
\gamma(\mathbf{z}) = \frac{P\, |h(\mathbf{z}, \boldsymbol{\Phi}_{\mathrm{PA}})|^2}{\sigma^2}.
\label{eq:snr}
\end{equation}

\subsection{Rate, Outage, and Secrecy Metrics}

Due to user location uncertainty, the SNR at Bob is a random variable. The instantaneous achievable rate is given by
\begin{equation}
R(\hat{\mathbf{u}}) = \log_2\!\left(1 + \gamma(\hat{\mathbf{u}})\right).
\label{eq:achievable_rate}
\end{equation}
To ensure quality of service (QoS), we impose an outage constraint requiring that the probability of the achievable rate falling below a target threshold $R_{\mathrm{th}}$ does not exceed $\epsilon$:
\begin{equation}
\mathbb{P}\!\left[R(\hat{\mathbf{u}}) < R_{\mathrm{th}}\right] \leq \epsilon.
\label{eq:outage_constraint}
\end{equation}

We further consider a potential Eve located at $\mathbf{e} = (x_e, y_e, 0)$, representing an estimated adversarial position (e.g., roadside or building edge). Eve’s received SNR follows the same propagation model:
\begin{equation}
\gamma(\mathbf{e}) = \frac{P\, |h(\mathbf{e}, \boldsymbol{\Phi}_{\mathrm{PA}})|^2}{\sigma^2}.
\label{eq:eve_snr}
\end{equation}
The instantaneous secrecy rate is defined as
\begin{equation}
R_s =
\Big[
\log_2\!\left(1 + \gamma(\hat{\mathbf{u}})\right)
-
\log_2\!\left(1 + \gamma(\mathbf{e})\right)
\Big]^+,
\label{eq:secrecy_rate}
\end{equation}
where $[x]^+ \triangleq \max(x,0)$. This metric quantifies the information-theoretic confidentiality of the transmission.

\subsection{Motivation for Adaptive Control under Uncertainty}

The secrecy-constrained optimization problem in \eqref{eq:optimization_problem} must be solved repeatedly as user locations, uncertainty levels, and channel conditions evolve over time. While the problem structure remains unchanged, the optimal antenna position $x_{\mathrm{PA}}$ and transmit power $P$ vary across scenarios due to stochastic user localization and adversarial dynamics. Solving this non-convex problem from scratch for each new realization is computationally prohibitive and incompatible with the stringent latency constraints imposed by mobility and channel coherence. In practice, the AP can rely only on a limited number of real-time observations, such as pilot-based SNR measurements and coarse user-location estimates, which further restrict the feasibility of iterative re-optimization. However, although the instantaneous solutions differ, all problem instances share a common physical and mathematical structure defined by the propagation model, outage formulation, and secrecy constraints. Exploiting this shared structure is key to enabling fast adaptation.

This observation motivates an adaptive control strategy that transfers knowledge across scenarios rather than treating each realization independently. Specifically, a few-shot, gradient-based meta-learning approach enables the AP to learn a transferable initialization offline, which can be efficiently refined online using only a small number of gradient updates. This allows the system to rapidly approximate the solution of \eqref{eq:optimization_problem} in real time while respecting both reliability and secrecy requirements.

\section{Secrecy-Constrained Optimization Problem Formulation}\label{sec_security_constraints} 
We now formulate the problem of jointly controlling the pinching-antenna position and the transmit power in a mathematically precise manner, while accounting for user-location uncertainty, secrecy requirements, and practical system constraints. As described in Section~\ref{sec_system_model}, the AP serves a legitimate user whose actual location lies within a circular uncertainty region centered at $\mathbf{u} = (x_u, y_u, 0)$. The pinching antenna location,
\begin{equation}
\boldsymbol{\Phi}_{\mathrm{PA}} = (x_{\mathrm{PA}}, 0, d),
\end{equation}
is \emph{adaptive} and can be repositioned along the waveguide. By adjusting $x_{\mathrm{PA}}$, the AP can reshape the propagation geometry to enhance the desired signal strength at the legitimate user while suppressing signal leakage toward a potential Eve.

\subsection{Reliability Constraint}
Due to uncertainty in the user’s true location $\hat{\mathbf{u}}$, the achievable rate becomes a random variable. To guarantee reliable communication, the system must ensure that the legitimate user attains a minimum target rate $R_{\mathrm{th}}$ with an outage probability no greater than $\epsilon$. This requirement is expressed as
\begin{equation}
\mathbb{P}\!\left[
\log_2\!\left(1 + \gamma(\hat{\mathbf{u}})\right) < R_{\mathrm{th}}
\right] \leq \epsilon,
\label{eq:robustness_constraint}
\end{equation}
which ensures that the target rate is achieved with probability at least $1-\epsilon$, despite user-location uncertainty.

\subsection{Secrecy Constraint}
In addition to reliability, the system must ensure confidentiality against a potential Eve located at a known or estimated position $\mathbf{e}$. To guarantee a sufficient advantage in channel quality for the legitimate user, we impose an average secrecy constraint given by
\begin{equation}
\mathbb{E}_{\hat{\mathbf{u}}}
\!\left[
\log_2\!\left(
\frac{1 + \gamma(\hat{\mathbf{u}})}
     {1 + \gamma(\mathbf{e})}
\right)
\right]
\geq R_{\mathrm{sec}},
\label{eq:secrecy_constraint}
\end{equation}
where $R_{\mathrm{sec}}$ denotes the minimum required secrecy rate. This constraint ensures information-theoretic confidentiality under uncertainty in the legitimate user’s location.

\subsection{Optimization Problem}

The optimization variables are the antenna’s horizontal position $x_{\mathrm{PA}} \in [0,L]$ and the transmit power $P \geq 0$. Since transmit power is a limited resource and increasing it benefits both the legitimate user and the Eve, the objective is to minimize $P$ while satisfying both reliability and secrecy requirements. The resulting optimization problem is formulated as follows:
\begin{equation}
\begin{aligned}
\min_{x_{\mathrm{PA}},\, P} \quad & P \\
\text{subject to} \quad
& \mathbb{P}\!\left[
\log_2\!\left(1 + \gamma(\hat{\mathbf{u}})\right) < R_{\mathrm{th}}
\right] \leq \epsilon, \\
& \mathbb{E}_{\hat{\mathbf{u}}}
\!\left[
\log_2\!\left(
\frac{1 + \gamma(\hat{\mathbf{u}})}
     {1 + \gamma(\mathbf{e})}
\right)
\right]
\geq R_{\mathrm{sec}}, \\
& x_{\mathrm{PA}} \in [0,L], \quad P \geq 0.
\end{aligned}
\label{eq:optimization_problem}
\end{equation}

This optimization problem is inherently non-convex due to the nonlinear channel model, the probabilistic outage constraint, and the secrecy-rate coupling between the legitimate user and the Eve. Consequently, it cannot be solved directly using standard convex optimization techniques. Moreover, the formulation implicitly assumes that a new solution can be computed whenever the user position or channel conditions change. In fast-varying mobile environments, such recomputation is computationally infeasible due to stringent latency constraints, motivating the need for fast and adaptive control mechanisms.

\subsection{Meta-Learning as an Adaptive Control Mechanism}
To address these challenges, we adopt a gradient-based meta-learning approach. By training a meta-learner over a distribution of prior tasks, each corresponding to different user and Eve locations and uncertainty levels, the system learns a transferable initialization of the control variables $(x_{\mathrm{PA}}, P)$. During deployment, this meta-learned initialization can be rapidly adapted to new scenarios using only a few pilot measurements and a small number of gradient updates, enabling real-time satisfaction of outage and secrecy constraints while minimizing transmit power.

This motivation leads to the gradient-based meta-learning framework presented in the next subsection. In particular, we develop a MAML-based controller as the primary solution due to its strong adaptation accuracy, while also considering a Reptile-based first-order meta-learning strategy as a lower-complexity baseline for comparison. These approaches enable few-shot adaptation and real-time decision-making in dynamic and adversarial wireless environments.

\section{Proposed Gradient-Based Meta-Learning Framework}\label{sec_proposed_framework}
Gradient-based meta-learning aims to learn model parameters that generalize across tasks and can be rapidly adapted to previously unseen tasks using only a few gradient updates. In this context, each task corresponds to a specific configuration of legitimate-user and Eve locations, user-location uncertainty, and the associated propagation geometry. By training an antenna-control policy over a distribution of such tasks, the AP can adapt to new environments using a limited number of pilot-based measurements and a small number of gradient updates.

Let $\mathcal{T} \sim p(\mathcal{T})$ denote a task drawn from the training distribution. Each task includes: (i) the estimated user location $\mathbf{u}$, (ii) the uncertainty radius $r_u$, (iii) the Eve location $\mathbf{e}$, and (iv) the corresponding channel environment. Let $\boldsymbol{\theta} \in \mathbb{R}^d$ denote the vector of learnable parameters. These parameters may correspond to a neural network that outputs the control variables $(x_{\mathrm{PA}}, P)$ from observed features, or to a differentiable policy that maps few-shot measurements to control actions.

\subsection{Task Loss Design}
Given a task $\mathcal{T}$, we define a task-specific loss as
\begin{equation}
\mathcal{L}_{\mathcal{T}}(\boldsymbol{\theta})
=
\mathbb{E}_{\hat{\mathbf{u}} \sim \mathcal{U}(\mathbf{u}, r_u)}
\!\left[
\mathcal{L}_{\mathrm{out}}(\hat{\mathbf{u}};\boldsymbol{\theta})
+ \lambda\, \mathcal{L}_{\mathrm{sec}}(\hat{\mathbf{u}},\mathbf{e};\boldsymbol{\theta})
\right],
\label{eq:task_loss}
\end{equation}
where $\mathcal{L}_{\mathrm{out}}$ penalizes violation of the target-rate requirement and $\mathcal{L}_{\mathrm{sec}}$ penalizes violation of the secrecy requirement. The parameter $\lambda$ controls the tradeoff between reliability and secrecy.

We adopt hinge-loss penalties \cite{Lee_2019_CVPR} that activate only when the corresponding constraints are violated. The outage loss is
\begin{equation}
\mathcal{L}_{\mathrm{out}}(\hat{\mathbf{u}};\boldsymbol{\theta})
=
\left[
R_{\mathrm{th}} - \log_2\!\left(1+\gamma(\hat{\mathbf{u}})\right)
\right]^+,
\label{eq:outage_loss}
\end{equation}
and the secrecy loss is
\begin{equation}
\mathcal{L}_{\mathrm{sec}}(\hat{\mathbf{u}},\mathbf{e};\boldsymbol{\theta})
\!=\!\!
\left[
R_{\mathrm{sec}}
\!-\!
\Big(
\log_2\!\left(1\!+\gamma(\hat{\mathbf{u}})\right)\!-\!\log_2\!\left(1\!+\gamma(\mathbf{e})\right)
\Big)
\right]^+.
\label{eq:secrecy_loss}
\end{equation}

\subsection{Meta-Training}
During meta-training, the inner-loop update simulates task-level adaptation. For task $\mathcal{T}_i$ and current parameters $\boldsymbol{\theta}$, the adapted parameters are obtained as
\begin{equation}
\boldsymbol{\theta}'_i
=
\boldsymbol{\theta}
-
\alpha \nabla_{\boldsymbol{\theta}}
\mathcal{L}_{\mathcal{T}_i}(\boldsymbol{\theta}),
\label{eq:inner_loop}
\end{equation}
where $\alpha>0$ denotes the inner-loop learning rate. 
The meta-objective seeks an initialization $\boldsymbol{\theta}$ that performs well after few-shot adaptation across tasks. This is captured by the meta-loss
\begin{equation}
\mathcal{L}_{\mathrm{meta}}(\boldsymbol{\theta})
=
\sum_{i=1}^{B}
\mathcal{L}_{\mathcal{T}_i}(\boldsymbol{\theta}'_i),
\label{eq:meta_loss}
\end{equation}
and the corresponding meta-update is
\begin{equation}
\boldsymbol{\theta}
\leftarrow
\boldsymbol{\theta}
-
\beta \nabla_{\boldsymbol{\theta}}
\mathcal{L}_{\mathrm{meta}}(\boldsymbol{\theta}),
\label{eq:meta_update}
\end{equation}
where $\beta>0$ is the meta-learning rate. This update promotes an initialization that is broadly adaptable across the task distribution.

The complete offline meta-training procedure is presented in Algorithm~\ref{alg:meta_training}.
\begin{algorithm}[h]
\caption{Meta-Training for Pinching-Antenna Control (MAML-Style)}
\label{alg:meta_training}
\SetAlgoLined
\KwIn{Task distribution $p(\mathcal{T})$, inner step size $\alpha$, meta step size $\beta$}
\KwOut{Meta-learned initialization $\boldsymbol{\theta}$}
Initialize $\boldsymbol{\theta}$ randomly\;
\While{not converged}{
    Sample batch $\{\mathcal{T}_i\}_{i=1}^B \sim p(\mathcal{T})$\;
    \ForEach{$\mathcal{T}_i$}{
        Evaluate $\mathcal{L}_{\mathcal{T}_i}(\boldsymbol{\theta})$ using \eqref{eq:task_loss}\;
        Compute $\boldsymbol{\theta}'_i \gets \boldsymbol{\theta} - \alpha \nabla_{\boldsymbol{\theta}} \mathcal{L}_{\mathcal{T}_i}(\boldsymbol{\theta})$\;
    }
    Compute $\mathcal{L}_{\mathrm{meta}} = \sum_{i=1}^{B} \mathcal{L}_{\mathcal{T}_i}(\boldsymbol{\theta}'_i)$\;
    Update $\boldsymbol{\theta} \gets \boldsymbol{\theta} - \beta \nabla_{\boldsymbol{\theta}} \mathcal{L}_{\mathrm{meta}}(\boldsymbol{\theta})$\;
}
\end{algorithm}

\subsection{Online Few-Shot Adaptation}
After offline meta-training, the model is deployed for real-time operation. When a new environment is encountered, the AP collects a small number of pilot-based measurements (e.g., received SNR/RSSI feedback and the location estimate $\mathbf{u}$) and performs a few gradient updates starting from the learned initialization $\boldsymbol{\theta}$. This enables rapid adaptation to previously unseen tasks and yields near-instantaneous decisions for antenna position and transmit power.
The corresponding online few-shot adaptation procedure executed during deployment is described in Algorithm~\ref{alg:online_adaptation}.
\begin{algorithm}[h]
\caption{Few-Shot Real-Time Adaptation (Online)}
\label{alg:online_adaptation}
\SetAlgoLined
\KwIn{New task $\mathcal{T}_{\mathrm{new}}$; meta-learned initialization $\boldsymbol{\theta}$; step size $\alpha$; adaptation steps $K$}
\KwOut{Adapted parameters $\boldsymbol{\theta}_{\mathrm{new}}$}
Receive pilot measurements and user estimate $\mathbf{u}$ with uncertainty radius $r_u$\;
Construct $\mathcal{L}_{\mathcal{T}_{\mathrm{new}}}(\boldsymbol{\theta})$ using \eqref{eq:task_loss}\;
\For{$t=1$ \KwTo $K$}{
    $\boldsymbol{\theta} \gets \boldsymbol{\theta} - \alpha \nabla_{\boldsymbol{\theta}} \mathcal{L}_{\mathcal{T}_{\mathrm{new}}}(\boldsymbol{\theta})$\;
}
Return $\boldsymbol{\theta}_{\mathrm{new}} \gets \boldsymbol{\theta}$\;
\end{algorithm}

This two-phase operation, offline meta-training followed by online few-shot adaptation, enables fast control without solving the underlying non-convex optimization problem from scratch for each new scenario. By leveraging transferable structure learned across tasks, the proposed approach supports low-latency adaptation in dynamic environments with user mobility and uncertainty.

\begin{figure}[!t]
	\centering
	\def\svgwidth{245pt}
	\fontsize{8}{4}\selectfont
	\scalebox{1}{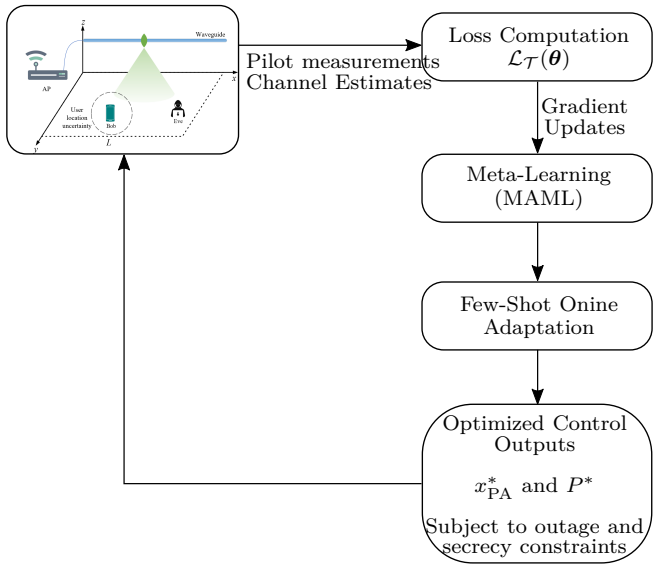}
	\caption{Proposed meta-learning-driven secure communication framework for waveguided pinching antenna systems.} \label{fig:flow_meta_learning}
\end{figure}

Fig.~\ref{fig:flow_meta_learning} illustrates the proposed meta-learning-driven framework for secure waveguided pinching-antenna systems. The AP serves the legitimate user (Bob) via the pinching antenna while Eve attempts to intercept the transmission. Pilot-based measurements are used to compute the task loss in \eqref{eq:task_loss}, enabling the policy to adapt its control decisions. The final output yields the adapted control variables $\{x_{\mathrm{PA}}^{*}, P^{*}\}$ that satisfy the reliability and secrecy requirements with low latency.

\section{Simulation Environment and Parameter Configuration}\label{sec_simulations}

To evaluate the proposed meta-learning based real-time antenna control framework, we conduct simulations for a mmWave downlink scenario with user-location uncertainty and the presence of a potential Eve. All simulations are implemented in MATLAB and repeated over $1000$ independent trials to ensure statistical reliability.
The simulation setup considers a waveguide of length $L=5$~m and antenna elevation $d=3$~m. The antenna position $x_{\mathrm{PA}} \in [0,L]$ is optimized in real time using the proposed meta-learned policy. The legitimate user lies within a circular uncertainty region centered at an estimated position $\mathbf{u}=(x_u,y_u,0)$ with uncertainty radius $r_u \in [0,2]$~m, capturing imperfections in positioning/tracking and mobility variations. The Eve position $\mathbf{e}$ is assumed to be known or estimated and remains fixed within each simulation instance.

The simulations adopt a carrier frequency $f_c=28$~GHz and bandwidth $B=100$~MHz. The transmit power is optimized over $P\in[0,1]$~W. The noise power is set as $\sigma^2 = N_0 B$ with noise power spectral density $N_0=-174$~dBm/Hz. The outage probability threshold is $\epsilon=0.05$, and the secrecy-rate requirement is $R_{\mathrm{sec}}=0.5$~bps/Hz. All parameters are listed in Table~\ref{tab:sys_params}.

\begin{table}[t]
\centering
\caption{System and Channel Parameters}
\label{tab:sys_params}
\begin{tabular}{l c}
\hline
\textbf{Parameter} & \textbf{Value} \\ \hline
Waveguide length, $L$ & $5~\mathrm{m}$ \\
Antenna height, $d$ & $3~\mathrm{m}$ \\
Carrier frequency, $f_c$ & $28~\mathrm{GHz}$ \\
System bandwidth, $B$ & $100~\mathrm{MHz}$ \\
Transmit power range, $P$ & $0$ to $1~\mathrm{W}$ \\
Noise PSD, $N_0$ & $-174~\mathrm{dBm/Hz}$ \\
Noise power, $\sigma^2$ & $N_0 B$ \\
User uncertainty radius, $r_u$ & $0$--$2~\mathrm{m}$ \\
Outage probability threshold, $\epsilon$ & $0.05$ \\
Target secrecy rate, $R_{\mathrm{sec}}$ & $0.5~\mathrm{bps/Hz}$ \\
Path-loss model & Free-space LoS \\
Monte Carlo trials & $1000$ \\
\hline
\end{tabular}
\end{table}

The meta-learning controller is trained using a MAML-style procedure over $2000$ randomized tasks and validated over $500$ unseen tasks. During online adaptation, the model performs $K$ inner-loop updates based on a small number of pilot samples. The key meta-training and adaptation parameters are summarized in Table~\ref{tab:algo_params}.

\begin{table}[t]
\centering
\caption{Meta-Learning and Adaptation Parameters}
\label{tab:algo_params}
\begin{tabular}{l c}
\hline
\textbf{Parameter} & \textbf{Value} \\ \hline
Meta-learning algorithm & MAML-style (gradient-based) \\
Task batch size, $B$ & $20$ \\
Inner-loop learning rate, $\alpha$ & $0.01$ \\
Meta-update learning rate, $\beta$ & $0.001$ \\
Adaptation steps, $K$ & $1$--$6$ \\
Training tasks & $2000$ \\
Validation tasks & $500$ \\
Pilot samples per task & $5$--$20$ \\
Tradeoff weight, $\lambda$ & $0.5$ \\
Latency budget (target) & $\leq 10$ ms \\
\hline
\end{tabular}
\end{table}



\subsection{Performance Results and Evaluation}
Performance is evaluated in terms of outage probability, secrecy rate, convergence latency, and energy efficiency. Each metric is assessed as a function of key system variables, including the user-location uncertainty radius, the number of adaptation steps, and the channel SNR. This evaluation procedure ensures both statistical robustness and practical relevance of the proposed methodology.

In this section, the results are organized into three components, each capturing a key performance dimension: (i) reliability and outage behavior, (ii) secrecy performance, and (iii) adaptation efficiency and resource utilization. This organization aligns with the dual objectives of the framework; namely, ensuring robust and secure communication under user-location uncertainty and achieving low-latency, energy-efficient adaptation in dynamic environments.
This categorization also serves three purposes. First, it enables focused analysis of individual performance metrics without conflating distinct behaviors; for instance, outage probability and secrecy rate, while related, are driven by different constraints in the optimization problem. Second, it follows a logical progression from physical-layer reliability to security robustness and finally to operational feasibility. Third, it provides a consistent basis for comparison with existing literature, where these metrics are typically reported separately.

For benchmarking, five baseline schemes are implemented under identical simulation conditions as follows:
\begin{itemize}
\item \textit{Reptile-based meta-learning}: A first-order meta-learning algorithm with lower computational complexity than MAML, illustrating the benefit of second-order meta-learning in highly dynamic settings.
\item \textit{Non-meta RL}: An RL agent trained from scratch for each new task, used to isolate the advantages of meta-initialization and knowledge transfer.
\item \textit{Conventional optimization}: A classical solution using full channel knowledge, serving as an idealized upper bound for performance.
\item \textit{Static antenna placement}: A non-adaptive configuration in which the pinching antenna remains fixed, highlighting the importance of spatial reconfiguration.
\item \textit{Power-only adaptation}: A heuristic that adjusts transmit power without repositioning the antenna, representing systems without spatial agility.
\end{itemize}


For fairness, all baseline schemes are implemented under identical channel models, uncertainty assumptions, and power constraints. The Reptile-based meta-learning baseline follows the same task formulation and loss functions as the proposed MAML framework, but employs first-order meta-updates without second-order gradients. The non-meta RL baseline is trained independently for each scenario without meta-initialization. The conventional optimization baseline assumes full channel knowledge and solves the secrecy-constrained problem via iterative numerical methods, serving as an idealized reference. Static antenna placement fixes the pinching antenna at a predetermined location, while the power-only adaptation baseline optimizes transmit power with the antenna position held constant.

\subsubsection{Reliability and Outage Performance}
This subsection evaluates system reliability through outage probability analysis, achievable rate distributions, and sensitivity to the user-location uncertainty radius $r_u$.

\begin{figure}[!t]
	\centering
    \includegraphics[width=0.4\textwidth]{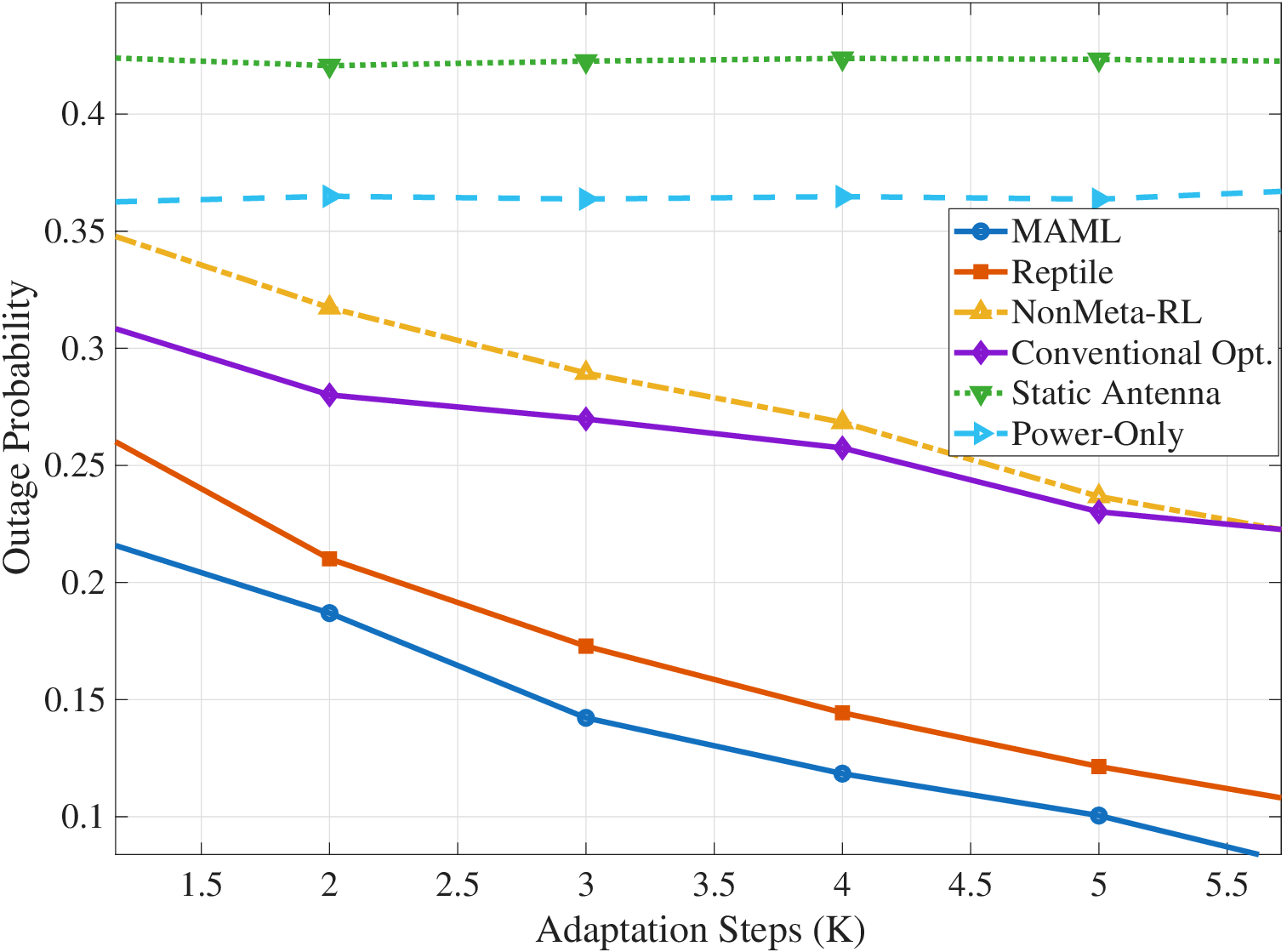}
	\caption{Comparison of outage probability performance as a function of adaptation steps $K$ for various approaches.}
	\label{fig3}
\end{figure}
Fig.~\ref{fig3} depicts the outage probability as a function of the number of adaptation steps $K$ for the proposed MAML-based framework and the five baseline schemes. The results show that the proposed method achieves the lowest outage probability across all values of $K$, demonstrating substantial gains even under a limited adaptation budget. For example, at $K = 3$, the outage probability of the MAML approach falls below $0.15$, outperforming Reptile by approximately $2\%$ and NonMeta-RL by more than $10\%$. The performance gap between MAML and Reptile expands with increasing $K$, highlighting the value of second-order gradient information in improving adaptation accuracy. 
Further, NonMeta-RL exhibits slower convergence due to the absence of meta-initialization, whereas the conventional optimization approach yields moderate performance but requires full CSI and incurs significantly higher computational latency, making it impractical for real-time operation. The static antenna and power-only baselines show negligible improvement with increasing K, underscoring the importance of spatial reconfiguration in reducing outage probability.
These results reveal that the proposed MAML-based adaptation  leverages a small number of gradient steps to achieve near-optimal reliability under mobility and CSI uncertainty.

\begin{figure}[!t]
	\centering
	\includegraphics[width=0.4\textwidth]{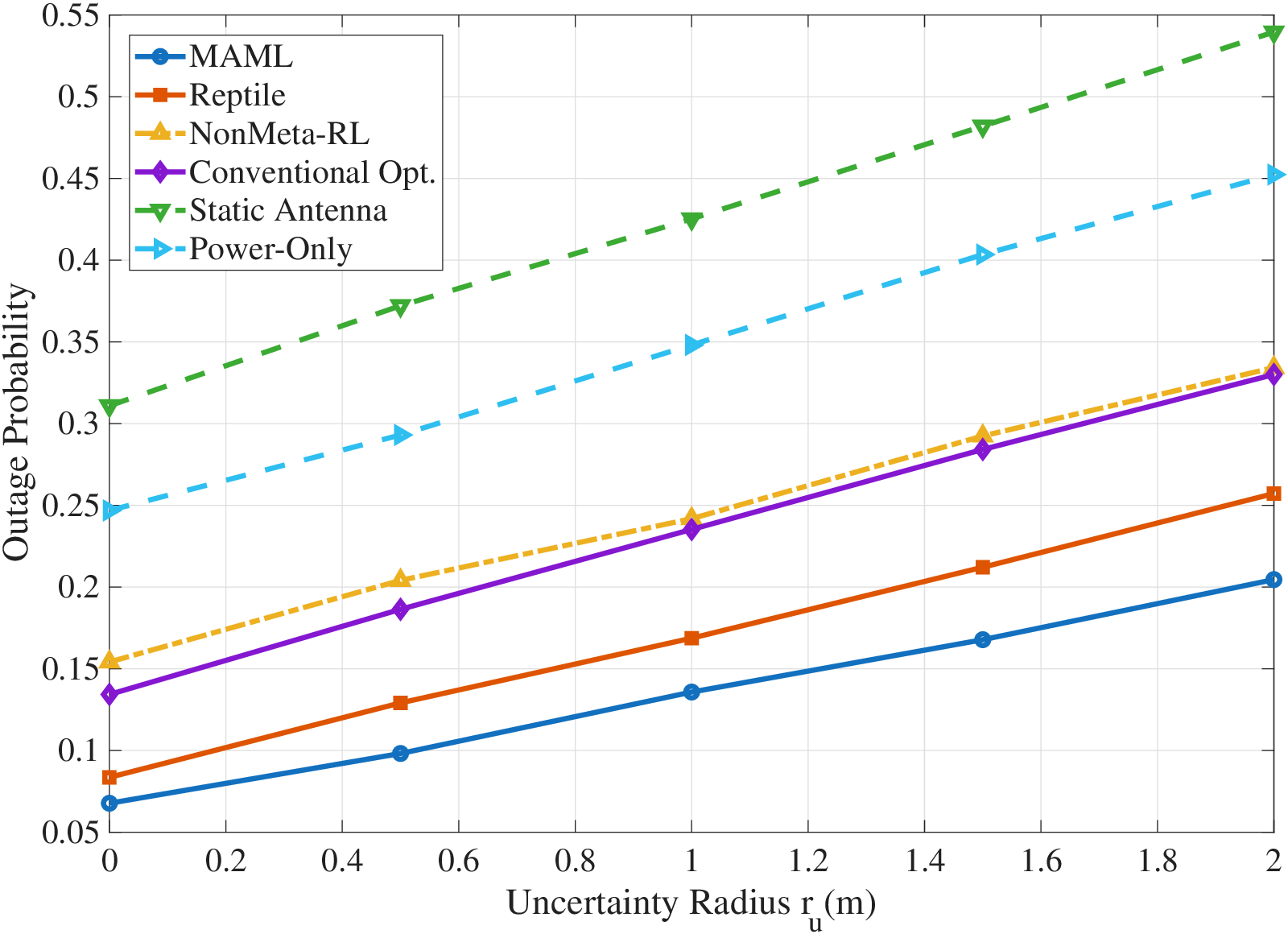}
	\caption{Outage probability performance as a function of user uncertainty radius $r_u$.}
	\label{fig4}
\end{figure}

Fig.~\ref{fig4} presents the outage probability as a function of the user uncertainty radius $r_u$ for the proposed MAML-based approach and the five baseline methods. Clearly, the outage probability increases monotonically with $r_u$ for all schemes due to reduced effective channel knowledge and a higher likelihood of rate violations within the uncertainty region. The proposed MAML framework consistently achieves the lowest outage probability across the entire range of $r_u$. At $r_u = 2\,\mathrm{m}$, MAML attains an outage probability of approximately $0.20$, representing a reduction of about $5\%$ compared with Reptile and more than $12\%$ compared with NonMeta-RL and conventional optimization.
Specifically, Reptile shows a similar trend but with slightly higher outage levels, reflecting the limitations of first-order updates relative to the second-order gradient information exploited by MAML. NonMeta-RL and conventional optimization perform comparably, both showing reduced robustness under high uncertainty,  NonMeta-RL due to slow adaptation and conventional optimization due to its reliance on prefect CSI. The static antenna and power-only control strategies yield higher outage probabilities, with the static configuration exceeding $0.5$ at $r_u = 2\,\mathrm{m}$, emphasizing their inadequacy in mobile or uncertain environments. 
These results confirm that the proposed framework effectively mitigates the impact of mobility-induced errors and imperfect location information, maintaining superior outage performance even under severe uncertainty conditions.

\begin{figure}[!t]
	\centering
	\includegraphics[width=0.4\textwidth]{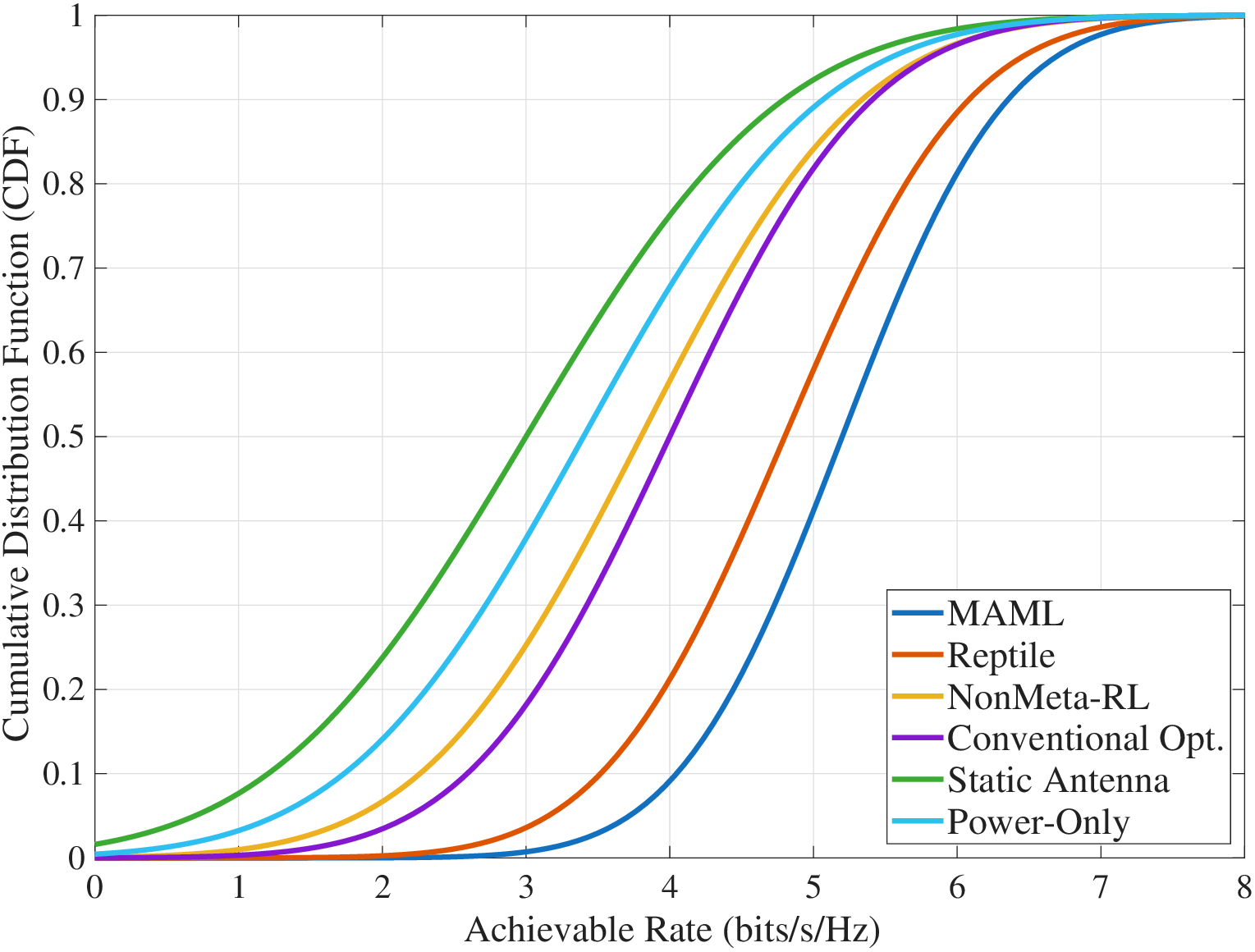}
    \caption{Cumulative distribution function (CDF) of the achievable rate for the proposed meta-learning-based framework and baseline schemes.}
	\label{fig5}
\end{figure}
Fig.~\ref{fig5} presents the cumulative distribution function (CDF) of the achievable rate for the proposed MAML-based method and the five baseline approaches. The results clearly show that the proposed MAML framework achieves the highest rates. The $50$th-percentile rate for MAML exceeds $4.8$~bits/s/Hz, whereas Reptile achieves approximately $4.4$~bits/s/Hz, and both NonMeta-RL and conventional optimization fall below $4.0$~bits/s/Hz. This performance advantage reflects MAML's ability to optimally adjust the antenna position and transmit power under channel and location uncertainties. Reptile follows a similar trend but exhibits slight degradation due to its reliance on first-order approximations in meta-updates. NonMeta-RL and conventional optimization produce nearly identical CDFs, indicating comparable yet suboptimal behavior, with a steeper decline in achievable rates at higher percentiles. The static antenna and power-only strategies impose substantial limitations, yielding median rates near $3.0$~bits/s/Hz and concentrating probability mass in the low-rate region. These results demonstrate that incorporating spatial reconfiguration through gradient-based meta-learning  enhances achievable rates,  particularly in dynamic wireless environments.

\subsubsection{Secrecy Performance}
This subsection evaluates secrecy performance through secrecy-rate analysis, secrecy-outage behavior, and robustness against variations in the Eve’s channel conditions.

\begin{figure}[!t]
	\centering
	\includegraphics[width=0.4\textwidth]{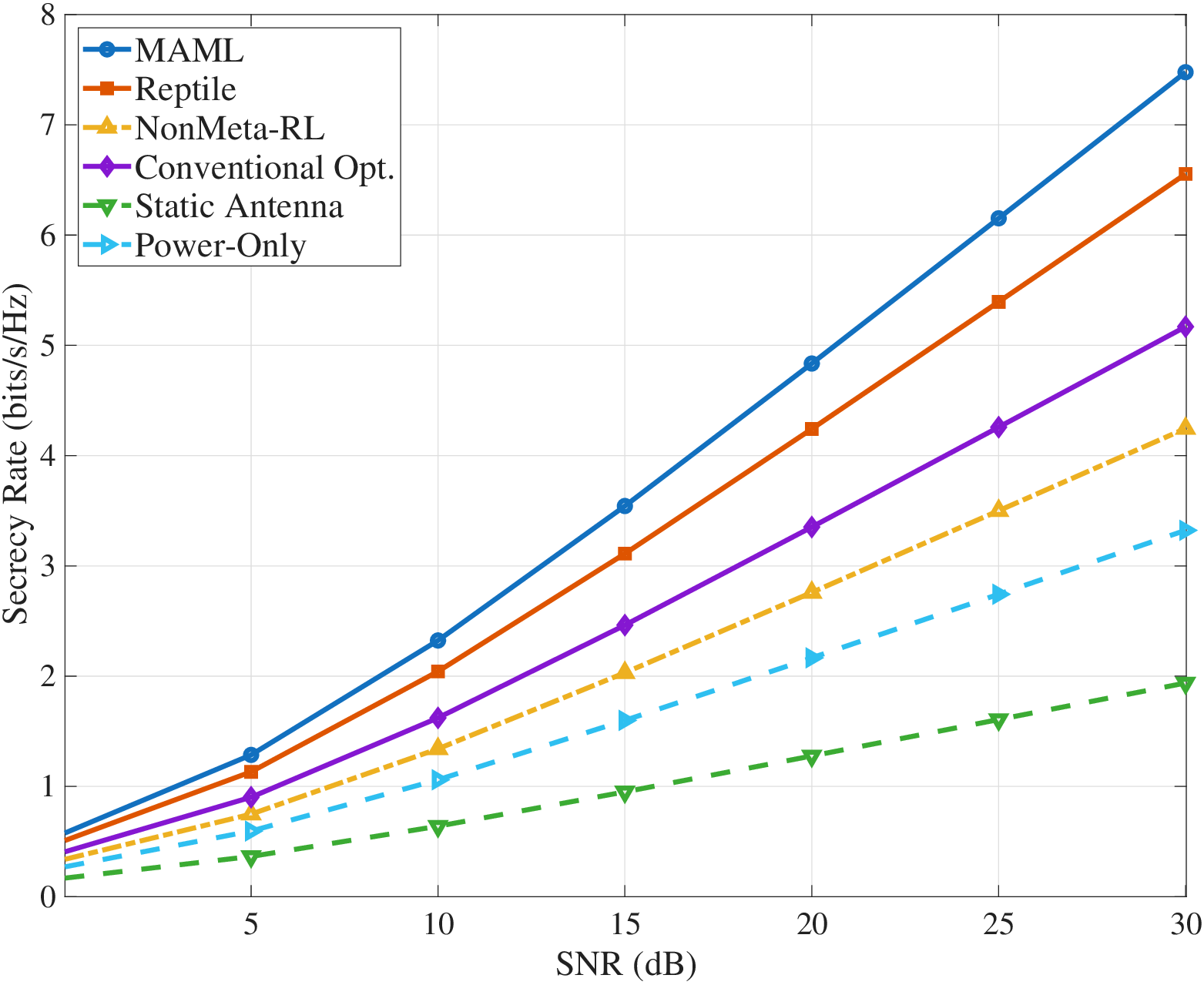}
	\caption{Secrecy rate versus SNR for the proposed meta-learning-based framework and baseline schemes.}
\label{fig6}
\end{figure}
Fig.~\ref{fig6} illustrates the secrecy rate as a function of the SNR for the proposed approach and the baseline schemes. The proposed method consistently achieves the highest secrecy rates across the entire SNR range, with the performance margin widening as SNR increases. At an SNR of $30$~dB, MAML attains a secrecy rate exceeding $7.4$~bits/s/Hz, outperforming Reptile and conventional optimization by approximately $0.8$~bits/s/Hz and $2.3$~bits/s/Hz, respectively. This improvement is due to MAML’s joint optimization of antenna positioning and transmit power, which enables rapid adaptation to fluctuations in both the legitimate and Eve channels. In addition, Reptile demonstrates a similar trend but with lower secrecy rates due to the reduced accuracy of its first-order meta-updates. Conventional optimization performs reasonably well under static or fully known channel conditions; however, its effectiveness diminishes in dynamic environments because it cannot exploit prior adaptation experience. NonMeta-RL shows lower secrecy performance, particularly at high SNR values, as it lacks the ability to leverage spatial reconfiguration efficiently. The power-only and static-antenna baselines produce the poorest results, with the static configuration remaining below $2.0$~bits/s/Hz even at $30$~dB, highlighting its vulnerability to eavesdropping under high-capacity conditions. These results show that  MAML can provide significant improvements in secrecy rate compared to both learning-based and traditional optimization methods, demonstrating its suitability for secure communication in dynamic environments with mobility and uncertainty.

\begin{figure}[!t]
	\centering
	\includegraphics[width=0.4\textwidth]{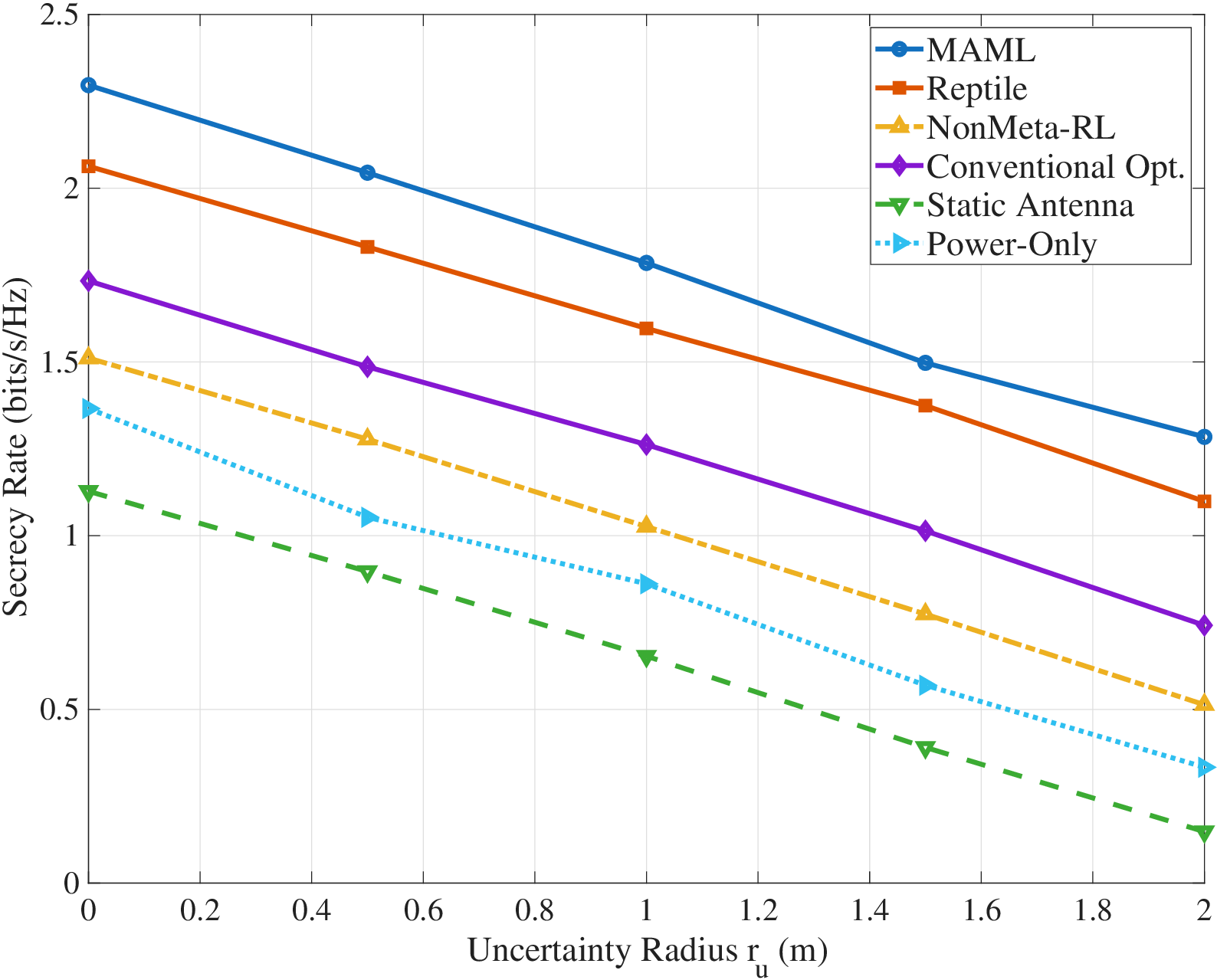}
	\caption{Secrecy rate performance as a function of user uncertainty radius $r_u$.}
	\label{fig7}
\end{figure}
Fig.~\ref{fig7} shows the secrecy rate as a function of the user uncertainty radius $r_u$ for the proposed MAML framework and the baseline approaches. Clearly, secrecy performance degrades for all schemes as $r_u$ increases, due to reduced beam targeting accuracy toward the legitimate user and an elevated likelihood of signal leakage to the Eve. The MAML-based approach  achieves the highest secrecy rates over all the  uncertainty range. At $r_u = 0$, MAML attains a secrecy rate of approximately $2.35$~bits/s/Hz, outperforming Reptile by roughly $0.25$~bits/s/Hz and conventional optimization by $0.6$~bits/s/Hz. Even at the maximum tested uncertainty radius of $2$~m, MAML maintains secrecy rates above $1.25$~bits/s/Hz, whereas the static antenna and power-only baselines fall below $0.4$~bits/s/Hz and $0.3$~bits/s/Hz, respectively.
Reptile provides the second-best performance but still below MAML due to its lower adaptation precision. Conventional optimization performs well in low-uncertainty scenarios but deteriorates rapidly as $r_u$ increases, highlighting its limited robustness to mobility-induced location errors. NonMeta-RL provides reduced secrecy rates at all uncertainty levels, underscoring the benefit of meta-initialization for few-shot adaptation. Meanwhile, the static and power-only schemes exhibit severe degradation, demonstrating their inadequacy for secrecy-critical applications in uncertain environments. 

\begin{figure}[!t]
	\centering
	\includegraphics[width=0.4\textwidth]{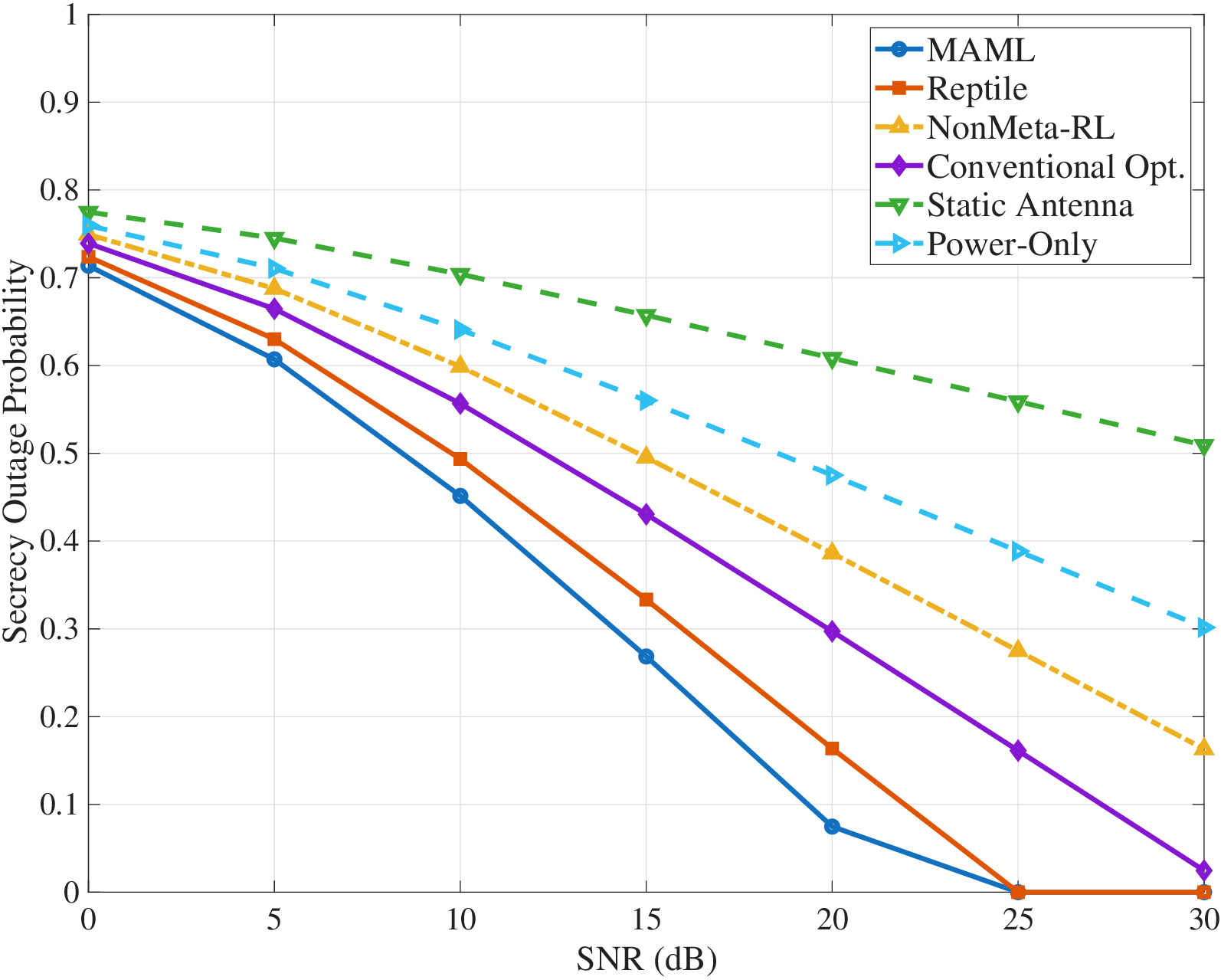}
\caption{Secrecy outage probability versus SNR for the proposed framework and baseline schemes.}
\label{fig8}
\end{figure}

Fig.~\ref{fig8} presents the secrecy outage probability as a function of the SNR. The MAML-based approach  achieves the lowest secrecy outage probability across the entire SNR range, with the performance gap widening in the medium-to-high SNR regime. At an SNR of $20$~dB, MAML reduces the secrecy outage probability to below $0.1$, whereas Reptile attains approximately $0.15$, and conventional optimization remains near $0.25$. In the low-SNR region ($\mathrm{SNR} < 5$~\textnormal{dB}), all adaptive methods exhibit similar performance since noise dominates over beamforming and adaptation gains. 
As SNR increases, the performance gains enabled by rapid, meta-learned spatial and power adaptation become more evident.
Reptile follows a similar downward trend but maintains higher outage levels, reflecting the limitations of its first-order meta-updates. Conventional optimization outperforms NonMeta-RL across all SNRs; however, both are surpassed by the meta-learning approaches due to their inability to leverage task-level knowledge transfer. NonMeta-RL demonstrates some adaptability but suffers from slower convergence and elevated outage probabilities, particularly for $\mathrm{SNR} > 15$~dB. The static antenna and power-only schemes exhibit negligible improvement with increasing SNR, maintaining outage probabilities above $0.5$ even at $30$~dB,  showing that they are ineffective for secure communications under uncertainty.

These results indicate that the proposed MAML framework  enhances secrecy performance and significantly improves the likelihood of meeting secrecy requirements across a broad range of SNR conditions, outperforming both traditional optimization and other learning-based baselines.

\subsubsection{Adaptation Efficiency and Resource Utilization} 
This subsection evaluates the adaptation efficiency and resource utilization of the proposed framework by analyzing convergence latency and the transmit power required to satisfy reliability constraints. These metrics assess the framework’s capability to operate under stringent real-time and energy-sensitive conditions.


\begin{figure}[!t]
	\centering
	\includegraphics[width=0.4\textwidth]{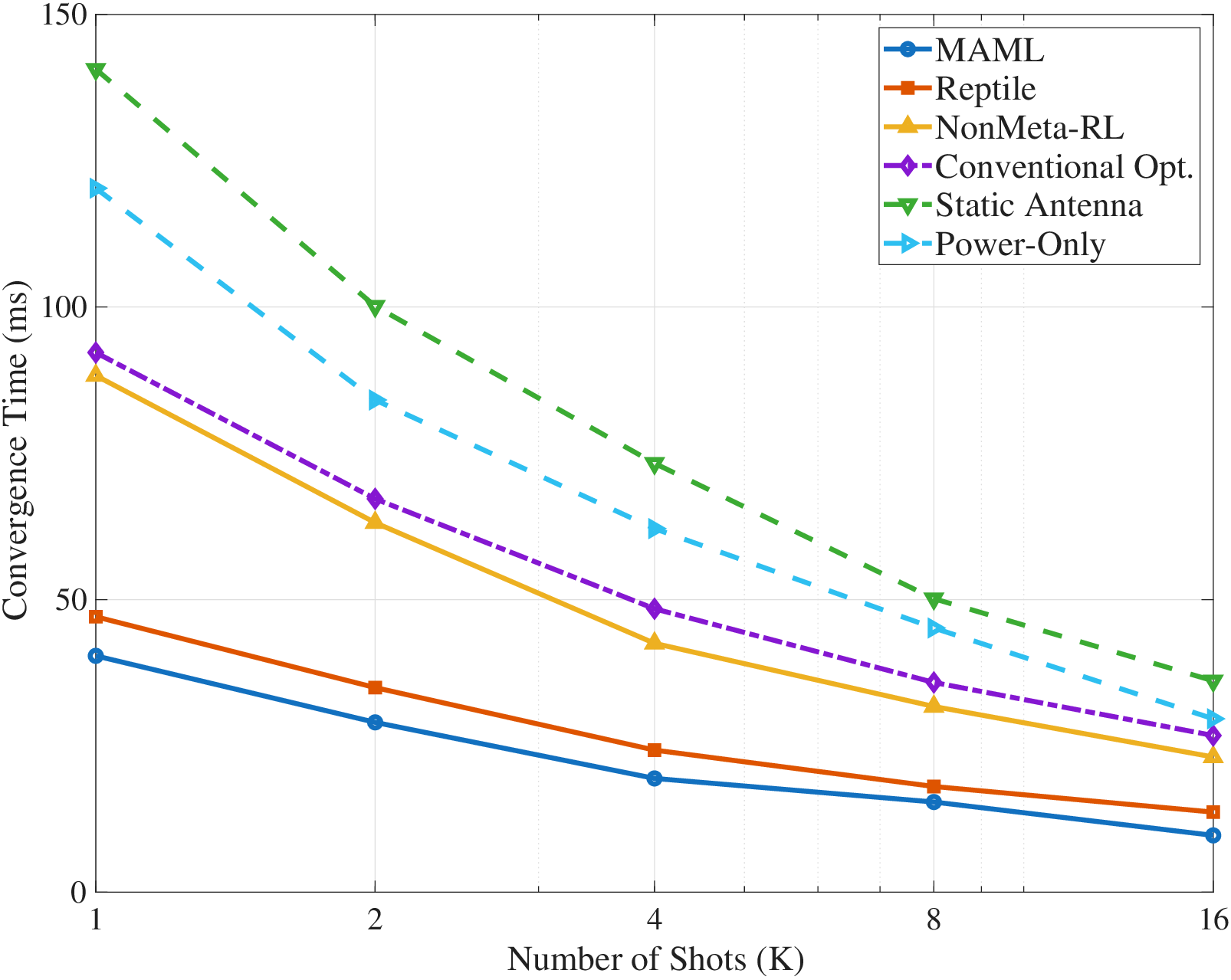}
\caption{Convergence time versus number of adaptation shots $K$ for the proposed framework and baseline schemes.}
\label{fig9}
\end{figure}
Fig.~\ref{fig9} shows the convergence time as a function of the number of adaptation shots $K$ for the proposed MAML-based approach and the baseline methods. MAML achieves the shortest convergence times across all values of $K$, demonstrating its ability to rapidly adapt to new tasks with minimal computational overhead. At $K = 1$, MAML converges in approximately $40$~ms, whereas Reptile requires about $47$~ms, and both conventional optimization and NonMeta-RL exceed $85$~ms. Convergence time decreases for all methods as $K$ increases due to the availability of additional adaptation data, with the most significant improvements observed for conventional optimization and NonMeta-RL, which begin with substantially higher initial latencies.
At $K = 16$, MAML maintains a convergence-time advantage of more than $10$~ms compared to the second-best baseline (Reptile), highlighting the benefits of superior meta-initialization and fewer gradient steps required for effective task-specific optimization. The static antenna and power-only schemes yield the longest convergence times, exceeding $130$~ms at $K = 1$ and remaining above $35$~ms even at $K = 16$, demonstrating their ineffectiveness for real-time adaptation in rapidly varying wireless channels.

\begin{figure}[!t]
	\centering
	\includegraphics[width=0.4\textwidth]{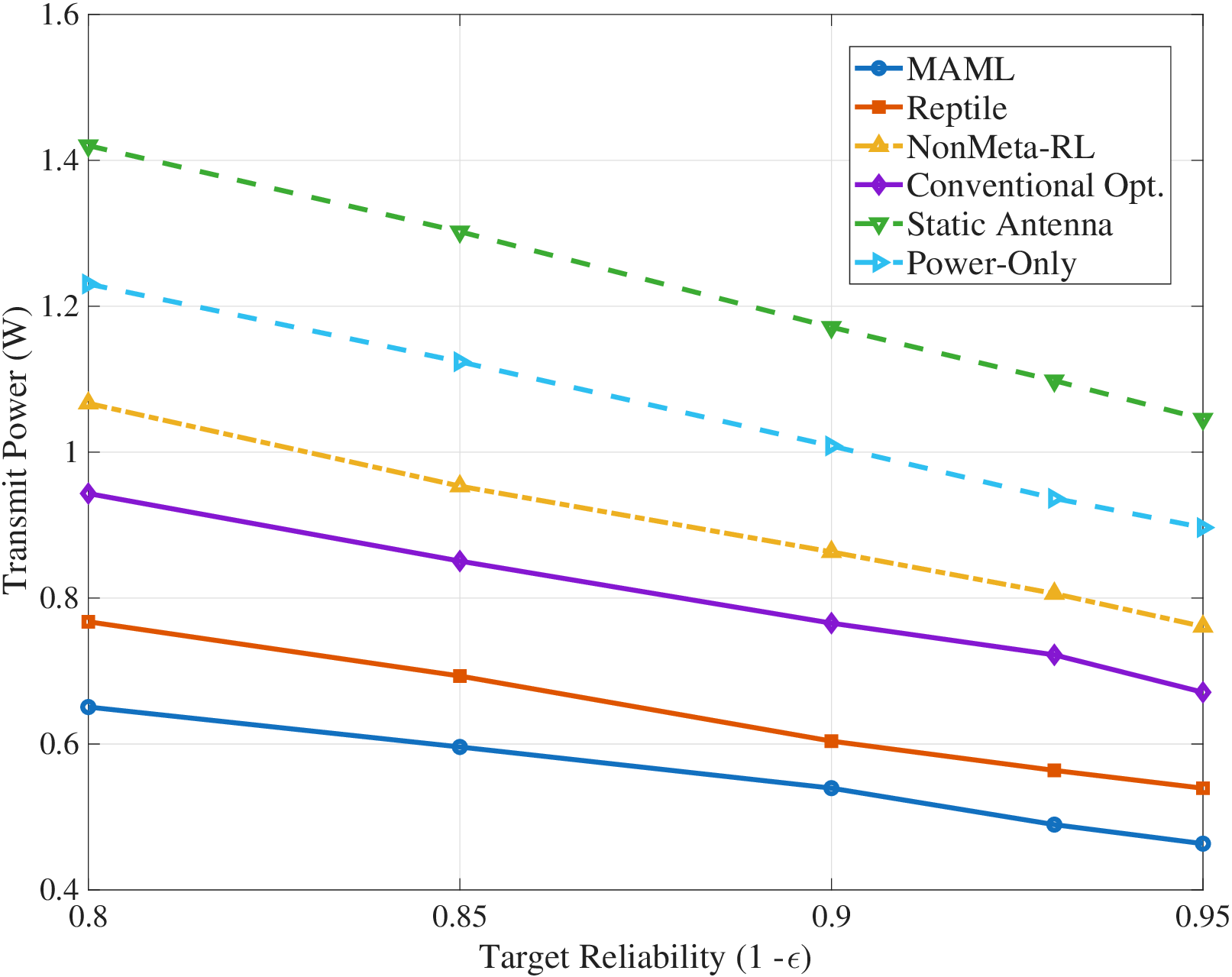}
\caption{Minimum transmit power versus target reliability $(1-\epsilon)$ for the proposed MAML-based framework and baseline schemes.}
\label{fig10}
\end{figure}
Figure~\ref{fig10} illustrates the relationship between the target reliability $(1-\epsilon)$ and the minimum transmit power required by the proposed MAML-based framework and the baseline schemes. Obviously, the MAML approach  achieves the lowest transmit power across the examined reliability range ($0.80$--$0.95$), demonstrating its effectiveness in satisfying reliability constraints while minimizing energy expenditure. For a reliability target of $0.95$, MAML requires only $0.48$~W, compared with $0.54$~W for Reptile, $0.68$~W for conventional optimization, and more than $1.0$~W for both the static-antenna and power-only baselines. The performance gap between adaptive meta-learning methods and static or partially adaptive schemes widens at higher reliability levels, emphasizing the importance of precise spatial and power adaptation for stringent reliability requirements.
Reptile exhibits slightly higher power consumption than MAML across all reliability levels due to its less accurate adaptation process. Conventional optimization performs competitively at moderate reliability levels but requires substantially higher power as $(1-\epsilon)$ approaches $0.95$, reflecting its inability to leverage task-level knowledge transfer. NonMeta-RL similarly requires greater transmit power than both meta-learning approaches owing to slower convergence and less efficient antenna-power configurations. The static-antenna and power-only baselines are the least efficient, requiring more than $1.2$~W even at the lowest reliability target, rendering them unsuitable for energy-constrained deployments.

\section{Conclusions}\label{sec_conclusions}
This paper introduced a gradient-based meta-learning framework for real-time control of waveguided pinching antennas in the presence of user-location uncertainty and adversarial eavesdropping. By leveraging the MAML paradigm, the proposed approach enables rapid few-shot adaptation of antenna positioning and transmit power without the need for full model retraining, thereby addressing the stringent latency requirements of dynamic wireless environments. 
A probabilistic system model was developed to characterize the outage and secrecy constraints induced by imperfect localization and mobility. Building on this model, a non-convex optimization problem was formulated, and a meta-learning solution was designed to approximate near-optimal adaptation policies efficiently. Extensive simulations demonstrated that the proposed MAML-based framework consistently outperforms competing approaches, including Reptile, NonMeta-RL, conventional optimization, static antenna placement, and power-only strategies, across key performance metrics such as outage probability, achievable rate distributions, secrecy rate, secrecy outage probability, and convergence latency. 

More importantly, the proposed method maintains robust performance even under severe location uncertainty, underscoring its practical value for real-world deployments.
In short, this solution advances pinching-antenna research beyond idealized static models and establishes a new direction for secure, adaptive, and low-latency physical-layer control in next-generation wireless systems. Future research directions include extending the framework to multi-antenna and multi-user settings, exploring distributed meta-learning for cooperative multi-AP architectures, and validating the proposed approach through hardware-in-the-loop experiments for integration into emerging 6G systems.

\bibliographystyle{IEEEtran}
\bibliography{IEEEabrv,References}

\end{document}